\documentclass{ws-procs9x6}

\usepackage{epsfig}

\begin{document}

\title{Nucleon Compton Scattering with Two Space--Like Photons$^1$}

\author{A. Afanasev} 

\address{Jefferson Lab, Newport News, VA 23606, USA\\
E-mail: afanas@jlab.org}

\author{I.~Akushevich$^{*}$}

\address{Duke University, Durham, NC 27708, USA\\
E-mail: aku@jlab.org}

\author{N.P.~Merenkov}

\address{NSC Kharkov Institute of Physics and Technology, Kharkov 63108, Ukraine\\
E-mail: merenkov@kipt.kharkov.ua}

\maketitle

\abstracts{
We calculated two--photon exchange effects for
elastic electron--proton scattering at high momentum transfers.
The corresponding nucleon Compton amplitude is defined by two space--like 
virtual photons that appear to have significant virtualities.  
We make predictions for a) a single--spin  beam asymmetry, and 
b) a single--spin target asymmetry or recoil proton polarization caused by an 
unpolarized electron beam. 
}

\footnotetext{$^1$ Talk presented by A. Afanasev at the Workshop on Exclusive Processes at High Momentum
Transfer, Jefferson Lab, Newport News, VA, May 15-18, 2002}
\footnotetext{$^{*}$ On leave from the National Center of Particle and High 
Energy Physics, 220040, Minsk, Belarus}

\section{Introduction}
The two-photon exchange mechanism in elastic electron-nucleon scattering 
can be observed experimentally by a) measuring the C-odd 
difference between electron--proton and positron--proton scattering  cross
sections;
b) analyzing deviations from the Rosenbluth formula and c) studying T-odd
parity--conserving single--spin observables. This paper concentrates on the latter.

Early measurements of the parity--conserving single--spin observables include
induced polarization of the recoil proton in elastic ep-scattering\cite{recoil} and 
the target asymmetry\cite{target}. These experiments were able to set only
upper bounds that appeared to be at one per cent level.
Corresponding theoretical calculations were given in Refs.\cite{th60s} and Ref.\cite{DeRujula71},
with deep--inelastic intermediate states considered in the latter.
  
\begin{figure}[ht]
\unitlength 1mm
\begin{center}
\begin{picture}(100,40)
\put(0,0){
\epsfxsize=10cm
\epsfbox{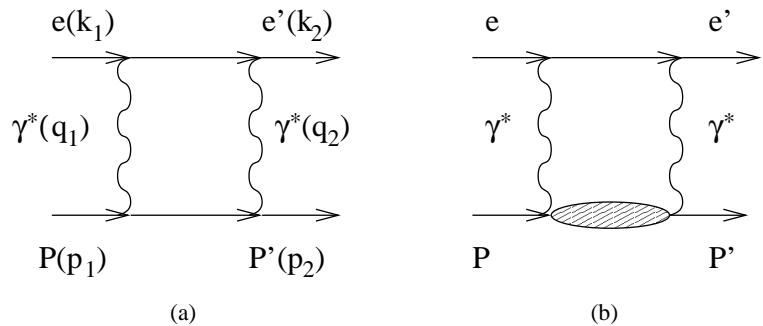}
}

\end{picture}
\end{center}
\caption{\label{2photon}
 Two-photon exchange mechanism responsible for single--spin asymmetries 
 in elastic {\it ep}--scattering. a) Elastic intermediate state.
 b) Inelastic intermediate states.}
\end{figure}

The transverse beam asymmetry of spin-$\frac{1}{2}$ particle scattering on 
a nuclear target was first calculated by N.F.~Mott\cite{Mott} in 1932, providing, 
for example, an operating principle for low--energy (about 1 MeV) electron beam
polarimeters \cite{MottPol}.  The measurement of this asymmetry at higher energies of several 
hundred MeV was reported recently by SAMPLE Collaboration \cite{Sample}. The observed magnitude 
of this effects is about $10^{-5}$ and it appears to be the only nonzero parity-conserving
single--spin effect measured so far for elastic ep--scattering.  
Here we present the first (to the best of our knowledge) 
theoretical evaluation of this asymmetry that takes nucleon structure into consideration. 
We also present results of our calculations of parity--conserving 
single--spin effects due to initial or final proton polarization in 
elastic $ep$--scattering. Since we are dealing with large transferred momenta, 
we describe excitation of intermediate hadronic states (Fig.\ref{2photon})
in terms of deep--inelastic structure functions of the non--forward
Compton amplitude with two space--like photons.  

\section{Formalism}

In QED, beam and target parity--conserving single--spin 
asymmetries and polarizations  are caused by 
the two--photon exchange mechanism (Fig.\ref{2photon}). 
In the leading order of electromagnetic coupling constant,
the imaginary (absorptive) part of the two--photon exchange 
amplitude interferes with a (real) amplitude
of the lowest--order one--photon exchange to produce a  
single--spin effect
due to transverse (namely, normal to the scattering plane) polarization of 
either the electron or the proton. As was noticed by De Rujula and collaborators
over 30 years ago\cite{DeRujula71}, the quantity which governs transverse polarization effects
is {\it the absorptive part of the non--forward Compton amplitude for off--shell
photons scattering from nucleons.} 

Using parity- and time-reversal invariance, one can demonstrate 
that a) the beam asymmetry is zero in the ultrarelativistic 
beam energies, b) beam and target asymmetries are independent 
observables and c) the target asymmetry and recoil proton
polarization are equal.

There are two basic contributions  to the two--photon exchange mechanism
shown in Fig.\ref{2photon} which differ 
by the intermediate hadronic state. In the first
case (Fig.\ref{2photon}a) the intermediate state is purely elastic, containing only a proton and electron. 
In the second case (Fig.\ref{2photon}b) the target proton is excited producing continuum 
of particles in the intermediate state. 

A general formula for the transverse single--spin asymmetries includes integration over
the loop 4--momentum $k$.
It can be written as\cite{DeRujula71}
\begin{equation}\label{Aelin}
A^{el,in}_{l,p}={8\alpha \over \pi^2} {Q^2\over D(Q^2) } \int
d W^2 {S+M^2-W^2 \over S+M^2}
{d Q^2_1 \over Q^2_1} {d Q^2_2 \over Q^2_2} {1\over \sqrt{K}}
\;
B^{el,in}_{l,p},
\end{equation}
where $S=2k_1 p_1$, $Q^2_{1,2}=-q^2_{1,2}$,  $M$ is the proton mass and notation for
particle momenta is shown in Fig.\ref{2photon}.
For the elastic intermediate state, the integration is two dimensional because of two
Dirac deltas that put the intermediate lepton and proton on the mass shell. 
It does not apply to the proton in the inelastic case, so the additional integration
over $W^2$ has to be done (from $M^2$ to $S+M^2$), resulting in a triple integral.
The quantity  $D(Q^2)$ comes from Born ($i.e.$, one--photon exchange) 
contribution. The formulae for the relevant
quantities read
\begin{eqnarray}
D(Q^2)&=&8[Q^4(F_1+F_2)^2+2S_m(F_1^2+\tau F_2^2)],
     \nonumber\\
S_m&=&S^2-S Q^2- M^2 Q^2, \tau=Q^2/4M^2, \\
     K&=&
\sqrt{1-z_1^2-z^2-z_2^2+2zz_1z_2}.\nonumber
\end{eqnarray}
Here $z$, $z_1$ and $z_2$ are cosines of the scattering angles in c.m.s.
They are related to $Q^2$, $Q_1^2$ and $Q_2^2$ as follows:
\begin{equation} \label{Q2vsTheta}
	 Q^2={S^2\over 2(S+M^2)}(1-z), \quad
	 Q^2_{1,2}={S(S-W^2+M^2)\over 2(S+M^2)}(1-z_{1,2}).
\end{equation}

\begin{figure}[ht]
\unitlength 1mm
\begin{center}
\begin{picture}(160,45)
\put(0,-5){
\epsfxsize=6cm
\epsfbox{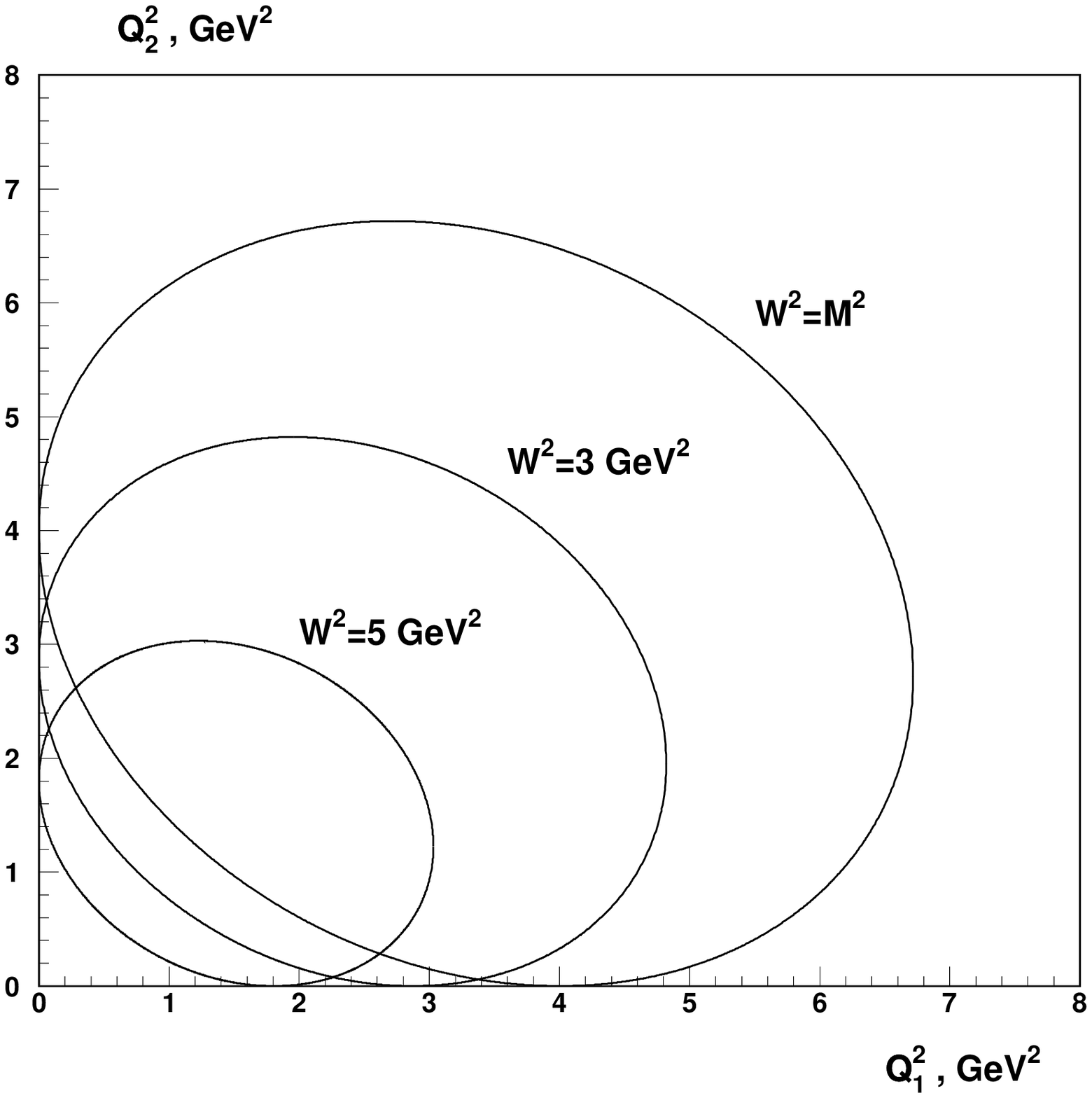}
}
\put(55,-5){
\epsfxsize=6cm
\epsfbox{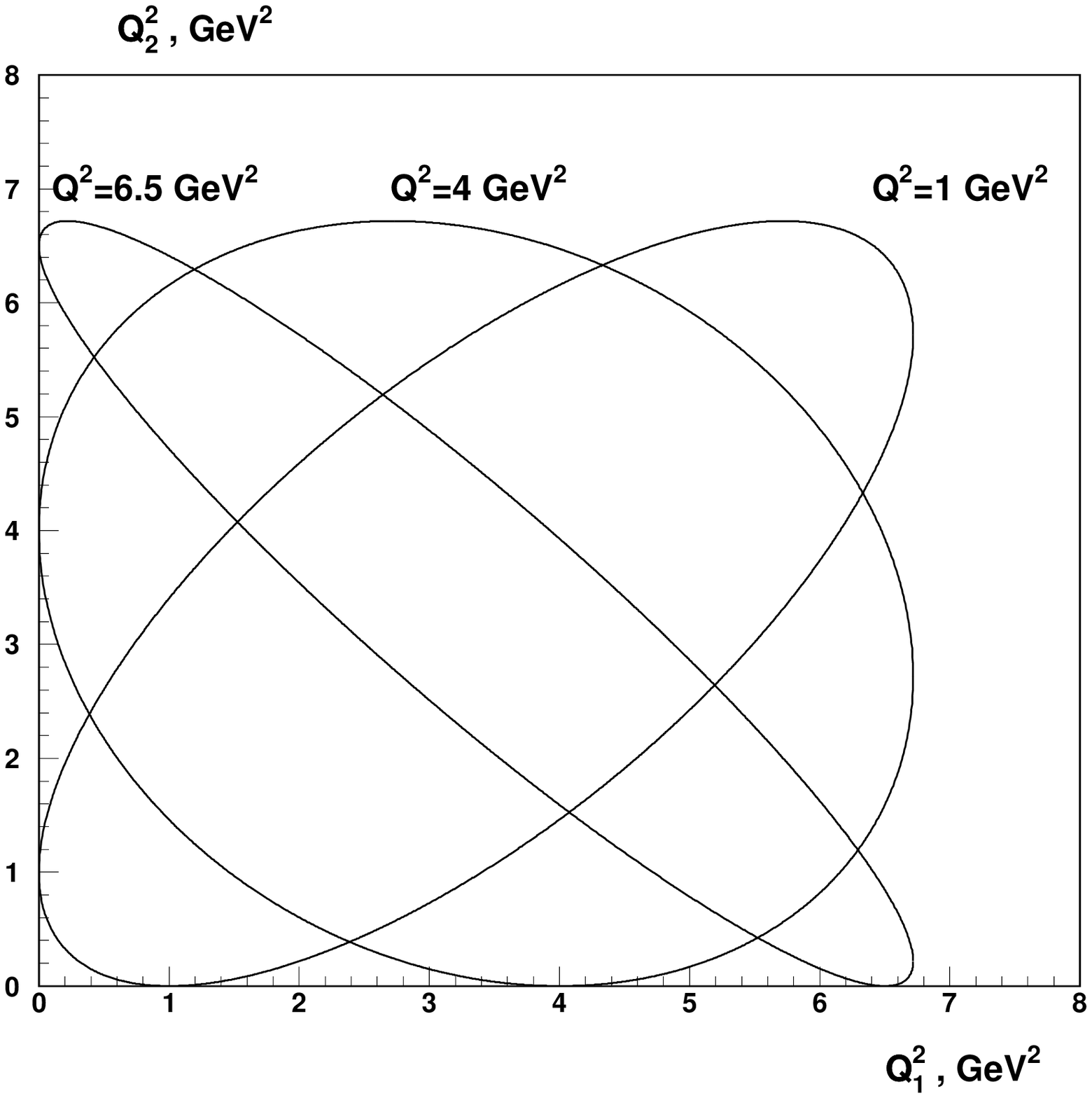}
}
\end{picture}
\end{center}
\caption{\label{intreg}
Integration region over $Q_1^2$ and $Q^2_2$ in Eq.(2) for elastic ($W^2=M^2$) and
inelastic contributions. The latter (left) is given for $Q^2$=4 GeV$^2$ and two values of $W^2$, which is
an integration variable in this case. The elastic case is shown on the right as a function
of external $Q^2$. The electron beam energy is $E_b$= 5 GeV.}
\end{figure}

The above formula (\ref{Q2vsTheta}) also sets the limits of the integration region for $Q_1^2$ and $Q^2_2$,
which is shown in Fig. \ref{intreg} for the representative electron beam energy $E_b=$ 5 GeV.
It can be seen from Fig. \ref{intreg}
that the virtualities of the exchanged photons, albeit limited, can become significantly
larger than the overall transferred momentum $Q^2$. Experimentally, by selecting the electron scattering angles
and beam energies, one can control the limits of photon virtualities contributing to the single--spin asymmetries.

The quantities $B$ in Eq.(\ref{Aelin}) result from contraction of leptonic and hadronic
tensors. Explicitly, they read
\begin{eqnarray}
B_l^{el,in}&=& \frac{i}{4}Tr[(\hat k_2+m)\gamma_\alpha (\hat
k+m)\gamma_\beta
\gamma_5 \hat\eta_l (\hat k_1+m) \gamma_\mu] \times
\nonumber\\&& \qquad
\frac{1}{4}Tr[(\hat p_2+M)W_{\alpha\beta}^{el,in}
(\hat p_1+M) \Gamma_\mu^*]
\\
B_p^{el,in}&=& \frac{1}{4}Tr[\hat k_2\gamma_\alpha \hat k
\gamma_\beta
\hat k_1 \gamma_\mu]
\frac{i}{4}Tr[(\hat p_2+M)\gamma_5\hat\eta_p W_{\alpha\beta}^{el,in}
(\hat p_1+M) \Gamma_\mu^*(q^2)],\nonumber
\end{eqnarray}
where $k(p)$ is the 4--momentum of the intermediate electron (proton),  
$m$ is the electron mass,  
$\eta_{l,p}$ is the electron (proton) polarization vector and
the nucleon electromagnetic current is parameterized in the form
\begin{equation}
\Gamma_\mu(q^2)=(F_1(q^2)+F_2(q^2)) \gamma_\mu -F_2(q^2)\frac{(p_1+p_2)_\mu} {2 M}.
\end{equation}

The only quantities than remain to be defined are elastic and inelastic
non--forward tensors $W_{\alpha\beta}$\footnote{Strictly speaking, we
deal only with imaginary parts of these tensors}. For the elastic case,
it is known exactly:
\begin{equation}
W_{\alpha\beta}^{el}=2\pi \delta(W^2-M^2) \Gamma_\alpha(q_2^2)(\hat
p+\hat q_1+M) \Gamma_\beta(q_1^2).
\end{equation}

To compute the non--forward inelastic hadronic tensor, we need model assumptions.
We use an expression for it inspired by Blumlein and Robaschik calculation \cite{BR},
\begin{equation}
W_{\alpha\beta}^{in}=2\pi
 \biggl((-g_{\alpha\beta}+{p^b_\alpha q_\beta^b+p^b_\beta q_\alpha^b
\over  q^bp^b})W_1+ p^b_\alpha p^b_\beta {W_2\over M^2} \biggr) {\hat q^b
\over q^bp^b}
\end{equation}
where $q^b=(q_1-q_2)/2$ and $p^b=(p_1+p_2)/2$.

These tensors are normalized such that in the forward
direction they reproduce conventional relations between inclusive structure
functions and elastic form factors,
\begin{equation}
W_1=2M\tau (F_1+ F_2)^2\delta(W^2-M^2),\ W_2=2M(F_1^2+\tau F^2_2)\delta(W^2-M^2).
\end{equation}

For the elastic intermediate state, we use unitarity to obtain model--independent 
analytic expressions for the target and beam asymmetries $B^{el}$,
\begin{eqnarray}\label{Bpl}
B_p^{el}&=&{\sqrt{S_m}\over 4M\sqrt{Q^2}}
\bigl[
Q^2_1Q^2_2 \bigl(S_q F_{1} (2 {\mathcal F}_{22}+{\mathcal F}_{+})-\frac{1}{2} (4
F_{1} M^2
- F_{2} Q^2)
(2 {\mathcal F}_{11}+{\mathcal F}_{+}) )
\nonumber\\&&
-S S_q Q^2 (2 F_{2} {\mathcal F}_{11}-F_{1} {\mathcal F}_{+}-F_{1} {\mathcal F}_{-})
+{(Q^2-Q^2_+) S\over 8 S_m} \sum_{ij}C^{p}_{ij}F_i{\mathcal F}_j \bigr)
\bigr]
\\
B_l^{el}&=&{m\sqrt{S_m}\over 4\sqrt{Q^2}}
\bigl[
(-2 Q^2_1Q^2_2 (F_{1}+F_{2}) (2 {\mathcal F}_{11}+{\mathcal F}_{+})+
{Q^2-Q^2_+\over 4 M^2 S_m} \sum_{ij}C^{l}_{ij}F_i{\mathcal F}_j \bigr)
\bigr]
\nonumber
\end{eqnarray}
where $Q_+^2=Q_1^2+Q^2_2,\ S_q=S+M^2-\frac{Q^2}{2}$ and expressions for the coefficients $C_{ij}$ are given in the 
Appendix.

The index $i=1,2$ corresponds to form factors $F_{1,2}=F_{1,2}(Q^2)$, 
and the index $j$ takes values $11,22,+,-$ where ${\mathcal F}_j$ are quadratic
combinations of elastic form factors  with the arguments $Q_{1,2}^2$:
\begin{eqnarray}
&&{\mathcal F}_{11}=F_1(Q_1^2)F_1(Q_2^2),\ 
{\mathcal F}_{22}=F_2(Q_1^2)F_2(Q_2^2),
\nonumber\\
&&{\mathcal F}_+=F_1(Q_1^2)F_2(Q_2^2)+F_1(Q_2^2)F_2(Q_1^2),
\\
&&
{\mathcal F}_-={Q_1^2-Q_2^2\over
Q^2}(F_1(Q_1^2)F_2(Q_2^2)-F_1(Q_2^2)F_2(Q_1^2)).\nonumber
\end{eqnarray}

The asymmetries $B^{el}_{l,p}$ are given in the symmetric form with respect to the transformation
$Q^2_1 \leftrightarrow Q^2_2$. It can be seen also 
that
$$ B_{l,p}^{el}(Q^2=Q^2_1)=
 B_{l,p}^{el}(Q^2=Q^2_2)=0
$$
so there are no infrared singularities in Eq.(\ref{Aelin}).

An inelastic contribution to the asymmetries reads
\begin{eqnarray}
B_{l}^{in}&=&{mQ^2W_1 \over 16 \nu_b N_s}\bigl[R_1+2R_2+8R_3\bigr]+
{mQ^2W_2 \over 32 M^2 \nu_b N_s}R_4
\nonumber\\
B_{p}^{in}&=&{(Q^2_+-Q^2)(SF_2-2M^2F_1)+2F_2S_qw_m
\over 128 \nu_b^2 N_s M^3}
(4M^2W_1T_1 + \\
&&\nu_bW_2T_2)Q^2\nonumber
\end{eqnarray}
where
\begin{eqnarray}
R_1&=& (2\nu_b(2S-2\nu_b-Q^2)F_1+(F_1+F_2)Q_+^2Q^2)
(w_mQ^2 - 2\nu_bS + \nonumber\\ 
&& 2M^2(Q^2-Q_+^2)),
\nonumber\\
R_2 &=& (w_mQ^2 - (Q^2-Q^2_+)S)\nu_bQ^2(F_1 + F_2),
\nonumber\\
R_3 &=& (S^2-M^2Q^2 - Q^2S)\nu_b^2F_1,
\nonumber\\
R_4&=& (\nu_b(4F_1M^2 -F_1Q^2 - 2F_2Q^2)+ 2(F_1+F_2)Q^2S_q)
(w_mQ^2-\nonumber\\ 
&&2\nu_bS+2M^2(Q^2-Q_+^2)),
\nonumber\\
T_1&=&2Q^2_+Sw_m+2Q^2(S-w_m)^2-2Q^2Q^2_+(S+M^2)- \\
&&Q^2Q_+^2(S-w_m)+ Q_+^4S,
\nonumber\\
T_2&=&4M^2(Q^2Q^2_+ - Q^2S_w - Q^2_+S)- 4Q^2_+S^2
+ Q^4S_w + \nonumber\\
&&3Q^2Q^2_+S + 8(S-Q^2)SS_w\nonumber,\\
w_m&=&W^2-M^2, \ \nu_b=w_m+\frac{Q_+^2-Q^2}{2}, \nonumber\\
N_s&=&\frac{1}{2} \sqrt{Q^2(S^2-M^2Q^2-SQ^2)}, \ S_w=S+M^2-W^2. \nonumber
\end{eqnarray}

In general, the structure functions $W_{1,2}$ are functions of four
invariant variables, $Q^2$, $Q^2_{1,2}$ and
$W^2$. These structure functions were neither measured nor calculated theoretically.
A possibility to construct a model for them was discussed in
Ref.\cite{DeRujula73}, where upper bounds were obtained from the positivity conditions.
Following Ref.\cite{DeRujula73}, we can write
\begin{equation}
W_{1,2}(W^2,Q^2_1,Q^2_2,Q^2)=
\left[W_{1,2}^{DIS}(W^2,Q^2_1)
W_{1,2}^{DIS}(W^2,Q^2_2) \right] ^{1/2}
F(Q^2),
\end{equation}
where we assumed additional form factor--like dependence $F(Q^2)$ on the overall 4--momentum transfer.
If the deep--inelastic conditions ($W>$2 GeV, $Q^2>$1 GeV$^2$) take place, 
the non--forward Compton form factor $F(Q^2)$ may be related to the integral
of Generalized Parton Distributions (GPD)\cite{GPD} at large transverse momenta $t$ (with the Mandelstam variable
$t$ equal to $-Q^2$ in our case). 
Note that since
the {\it absorptive} part of the non--forward Compton amplitude contributes, then $x=\xi$ 
part of GPDs is selected similar to the single--spin asymmetry arising from interference
of the Bethe--Heitler process with virtual Compton scattering\cite{Diehl}. The computed quantities are then described
by the zeroth moments of nucleon GPDs and it is therefore natural to assume for further
estimates that the introduced form factor $F(Q^2)$ depends on $Q^2$ like the nucleon
form factor described, with a good accuracy, by the dipole formula $F(Q^2)=(1+\frac{Q^2}{0.71 GeV^2})^{-2}$.

This model choice for the non--forward Compton amplitude has three main properties. 
It is symmetric with respect to the transformation
$Q^2_1 \leftrightarrow Q_2^2$, has a correct forward ($Q^2=0$) limit
and assumes form factor--like suppression with respect to the overall transferred momentum $Q^2$.
The latter was not considered in the early papers\cite{DeRujula71,DeRujula73}, leading to dramatic
overestimates of the two--photon--exchange effects for elastic $ep$--scattering. 

\section{Numerical results and conclusions}

\begin{figure}[ht]
\unitlength 1mm
\begin{center}
\begin{picture}(160,45)
\put(-5,-10){
\epsfxsize=6cm
\epsfbox{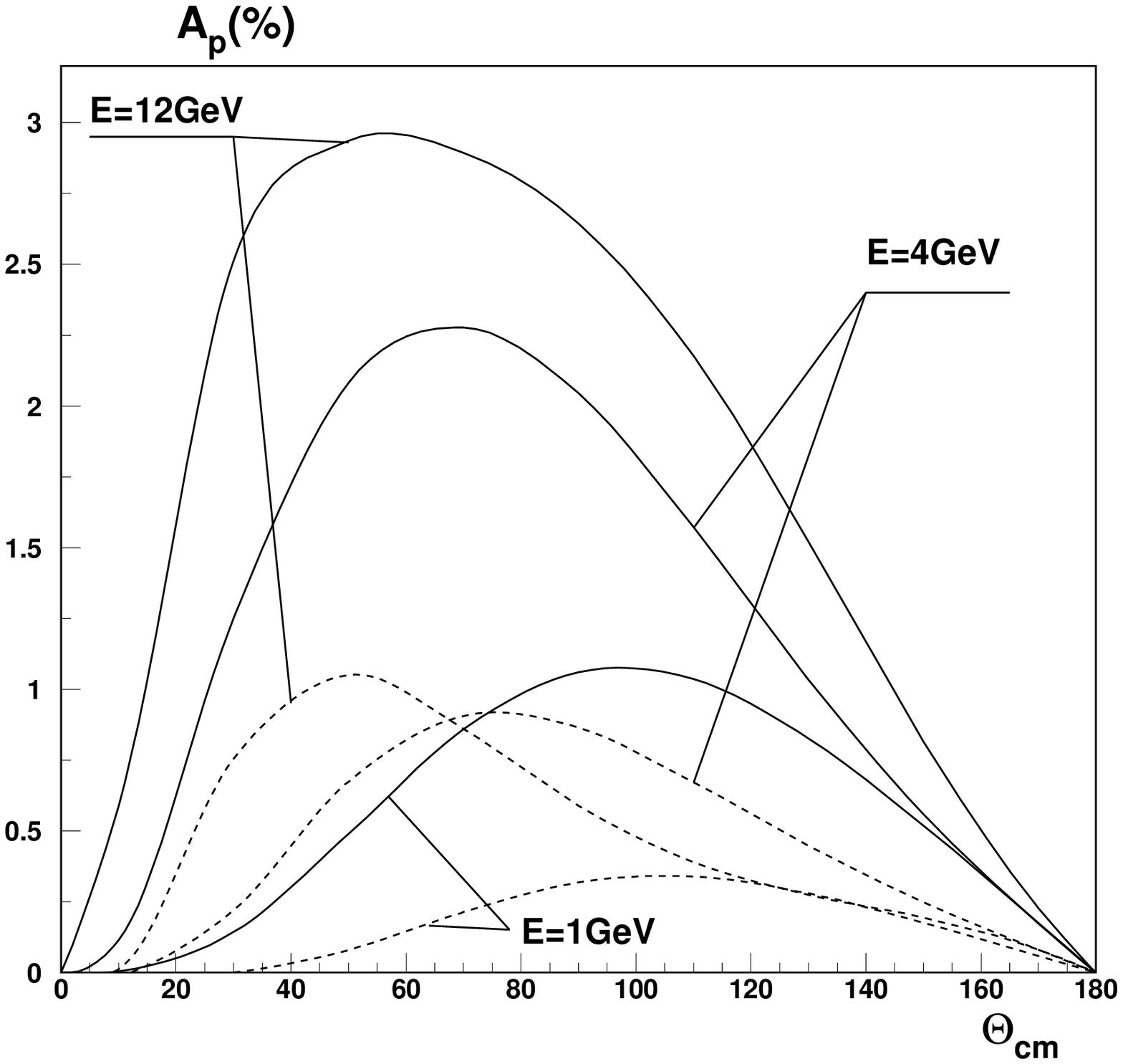}
}
\put(55,-10){
\epsfxsize=6cm
\epsfbox{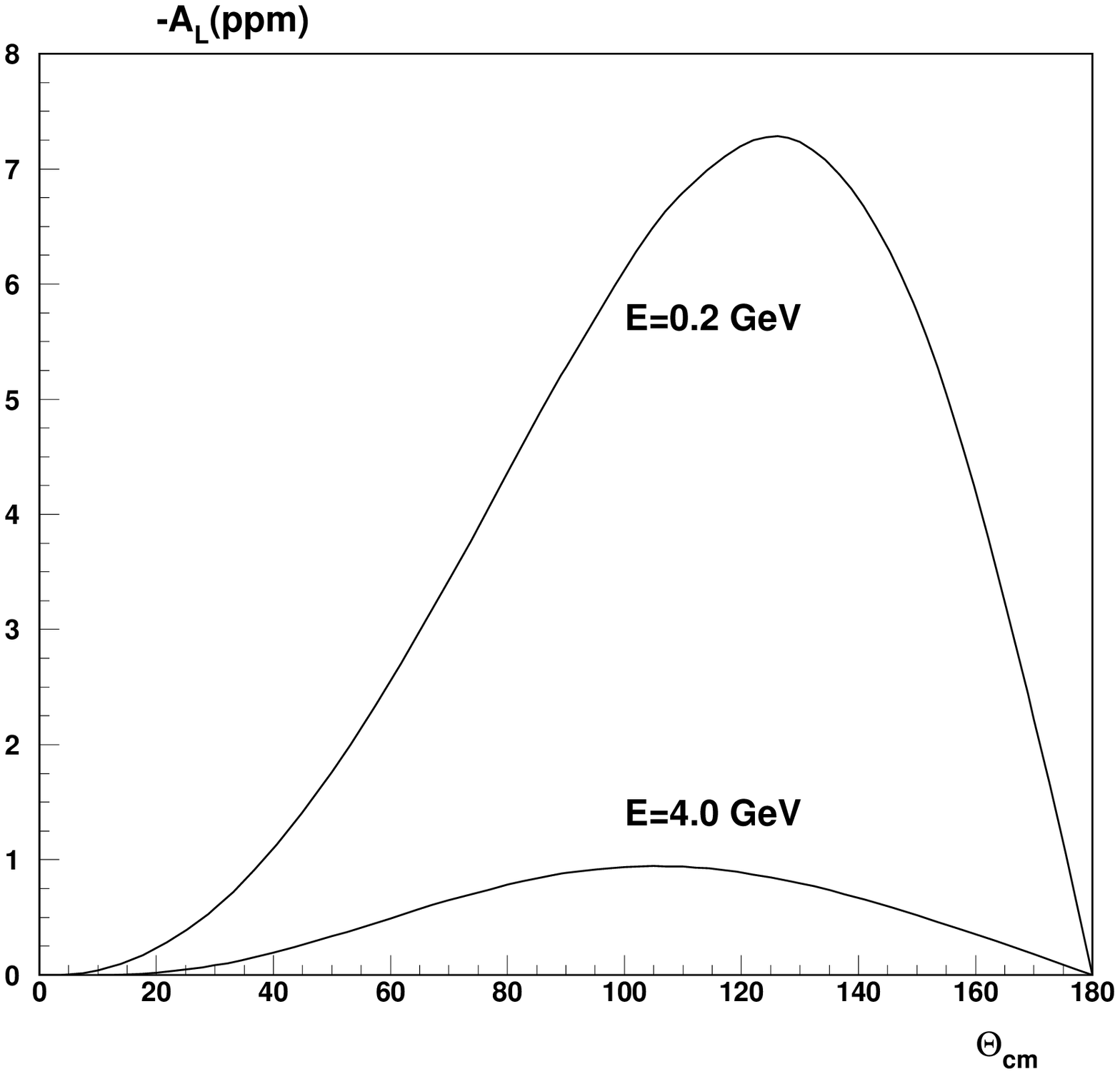}
}
\put(62,20){
\epsfxsize=2.2cm
\epsfysize=2.2cm
\epsfbox{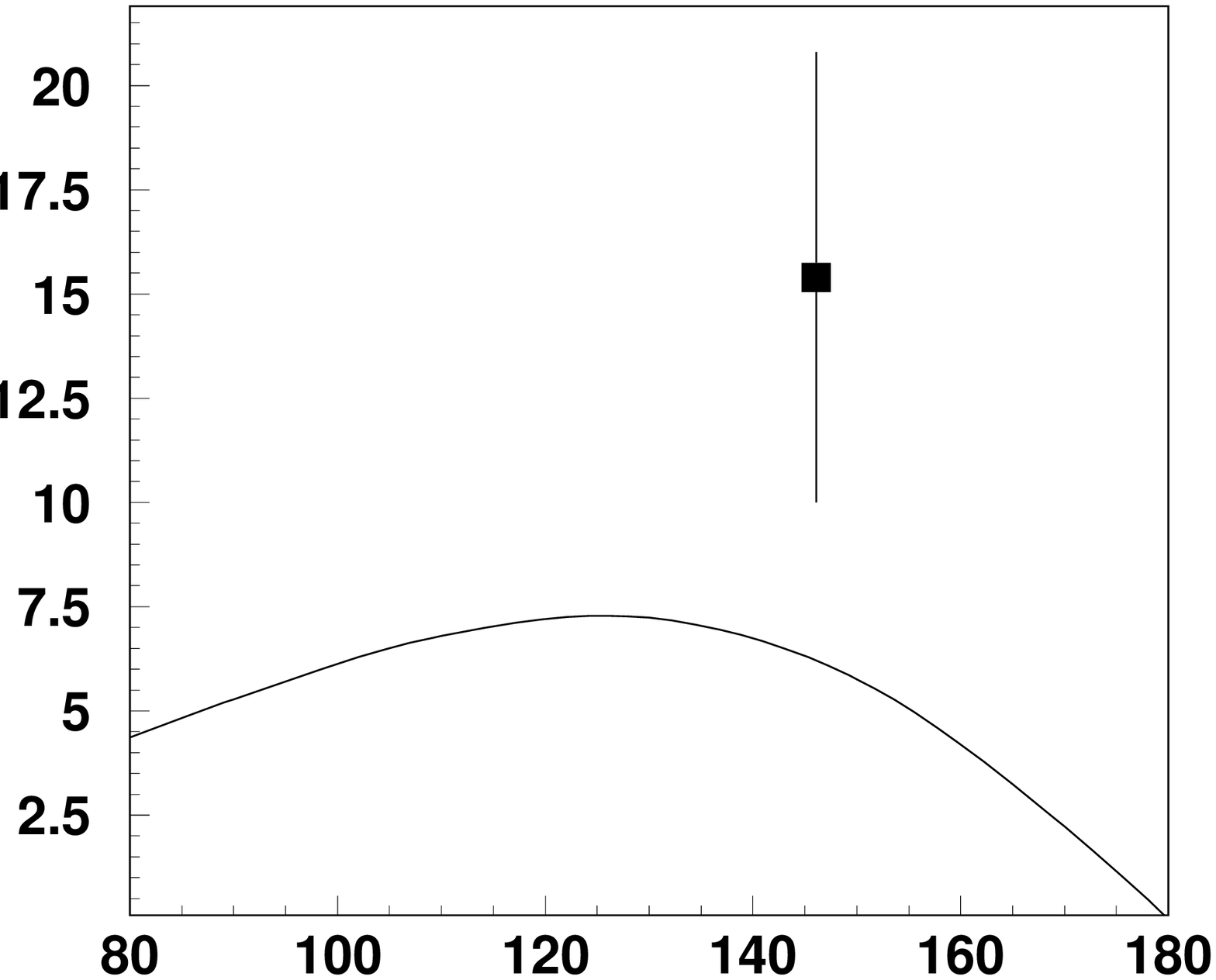}
}
\end{picture}
\end{center}
\caption{\label{mainres}
Proton (left) and lepton (right) asymmetries. Both elastic and total contribution are
shown for proton asymmetry and only elastic one for lepton asymmetry. The plot
in the insert gives comparison with SAMPLE data$^7$ at $E_B$= 0.2 GeV.}
\end{figure}

\begin{figure}[ht]
\unitlength 1mm
\begin{center}
\begin{picture}(160,45)
\put(0,-20){
\epsfxsize=6cm
\epsfbox{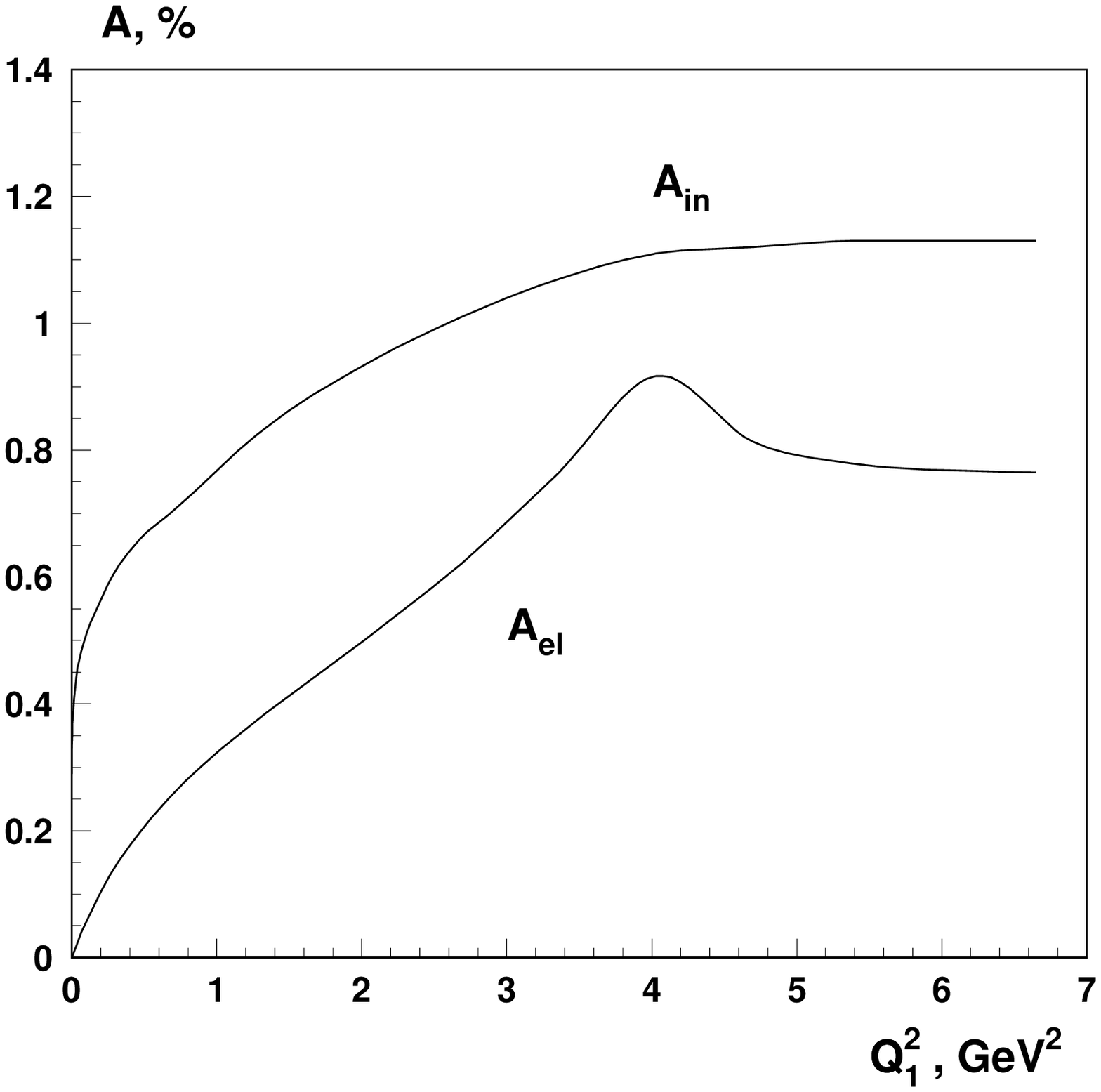}
}
\put(55,-20){
\epsfxsize=6cm
\epsfbox{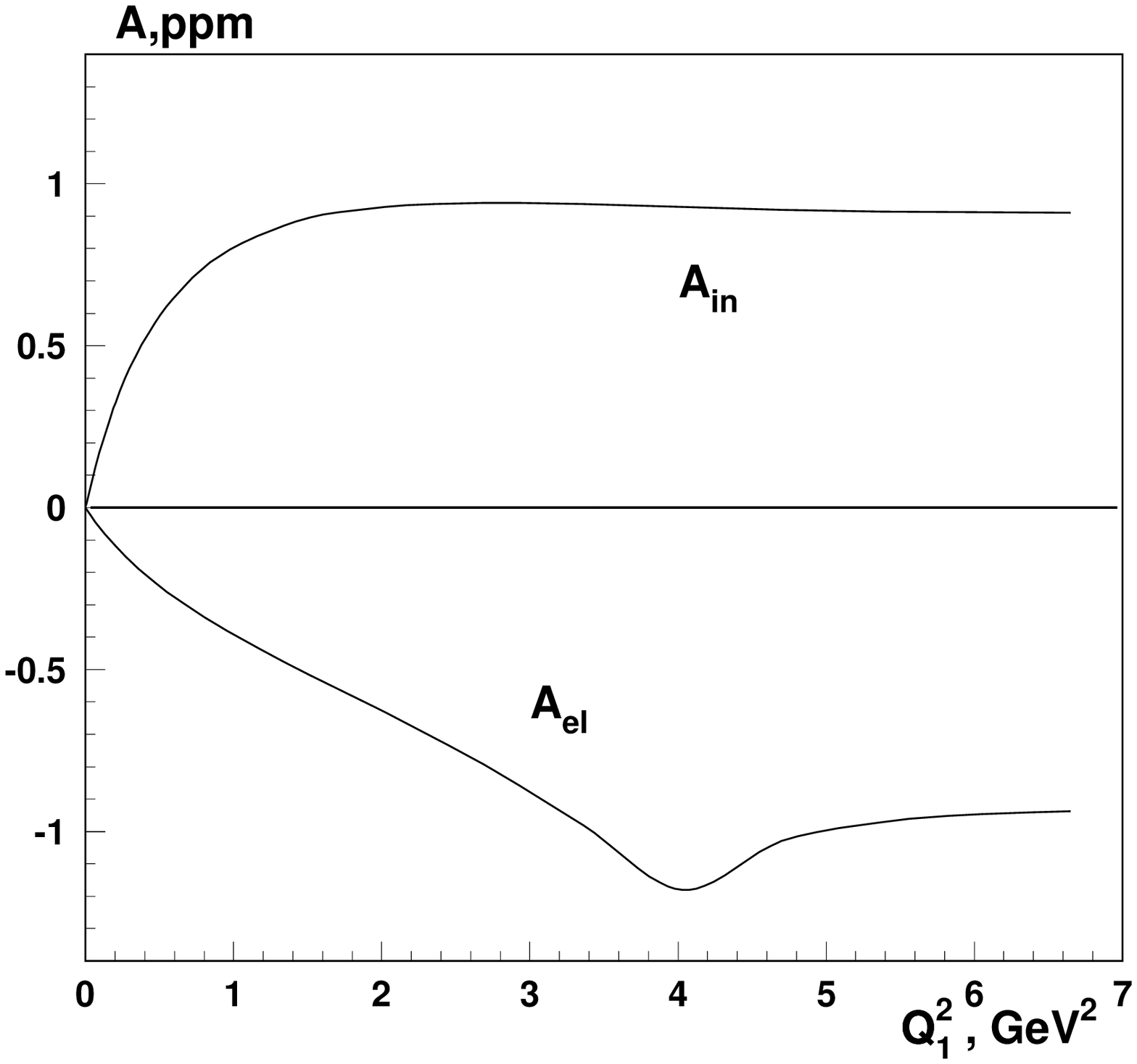}
}
\end{picture}
\end{center}
\caption{\label{lep}
Elastic and inelastic contributions to the target (left) and beam (right) asymmetries
versus the upper integration limit over the virtuality of one of the two exchanged photons.
Kinematics is as in Fig.2. Importance of high virtualities is evident.}
\end{figure}

Our results for the single--spin asymmetries are presented in Fig.\ref{mainres}. The asymmetries
are kinematically suppressed in the forward and backward directions by a factor $\sin(\Theta_{c.m.})$.
The target asymmetry increases with increasing beam energies, while the beam asymmetry decreases due to the
additional supression factor $m/E_b$. The target asymmetry (= recoil polarization) is 
evaluated at the per cent level. Below the pion threshold, only the nucleon intermediate 
state is allowed and the calculation becomes model--independent, 
based only on unitarity and known values of
proton form factors. At higher energies, the contribution from excited intermediate hadronic states
exceeds the elastic contribution. As can be seen from Fig.\ref{lep}, the integration region where
both exchanged photons are highly virtual plays an important role. 
Measurements of the single--spin asymmetries due to two--photon exchange provide
information about the absorptive part of the virtual Compton amplitude with two space--like
photons at large values of the Mandelstam variable $t$.

To our knowledge, this is the first published calculation of the
single--spin beam asymmetry in elastic $ep$--scattering 
for the kinematics where nucleon structure effects become important. 
Our results appear to be in reasonable agreement with recent SAMPLE data\cite{Sample}. 

\section*{Acknowledgements}

The work of IA and AA was supported by the U.S. Department of Energy under contract DE--AC05--84ER40150. 
We thank A.~Radyushkin, S.~Wells and S.~Brodsky for useful discussions.

\section*{Appendix}
\label{sec:AppA}
\setcounter{equation}{0}
\renewcommand{\theequation}{A.\arabic{equation}}
Expressions for the coefficients $C_{i,j}$ in Eq.(\ref{Bpl}) are as follows
\begin{eqnarray}
C^{p}_{1,11}&=&4 M^2 Q^2_+ (Q^2 + 2 S)
\nonumber\\
C^{p}_{1,22}&=&2 (-(Q^2-Q^2_+) (2 S^2-M^2 Q^2)+Q^4 S)
\nonumber\\
C^{p}_{1,+}&=& (4 Q^2 S_q (S+M^2)+Q^2_+ (4 M^2 Q^2 + 4 M^2 S - 2 S S_q))
\nonumber\\
C^{p}_{1,-}&=&-Q^2 (2 M^2 Q^2 - 2 S S_q)
\nonumber\\
C^{p}_{2,11}&=&- 4 Q^2 (2 S_q (S+M^2)+S Q^2_+)
\nonumber\\
C^{p}_{2,22}&=&-Q^2 (-(Q^2-Q^2_+) (S-M^2) - 2 S^2 + 2 Q^2 S) S/M^2
\nonumber\\
C^{p}_{2,+}&=& -3 Q^2 S Q^2_+
\nonumber\\
C^{p}_{2,-}&=&S Q^4
\nonumber\\
C^{l}_{1,11}&=&2 M^2 (2 Q^2_+ M^2 Q^2 + Q^2_+ Q^2 S + 2 Q^2_+ S^2+ 8 M^4Q^2  +12 M^2 Q^2 S)
\nonumber\\
C^{l}_{1,22}&=&Q^2  ((Q^2-Q^2_+) M^2 (S+2 M^2) - S^2 (Q^2_++ 4 M^2))
\nonumber\\
C^{l}_{1,+}&=&M^2  (4 Q^2_+ M^2 Q^2 + 3 Q^2_+ Q^2 S + 2 Q^2_+ S^2 + 2M^2Q^4 + Q^4 S - 6 Q^2 S^2)
\nonumber\\ 
C^{l}_{1,-}&=&-2 M^2 Q^4 (M^2 + S)
\nonumber\\
C^{l}_{2,11}&=&-4 M^2 (M^2 Q^4+Q^2 S^2+Q^4 S-Q^2_+ S^2)
\nonumber\\
C^{l}_{2,22}&=&-S^2 Q^2 (Q^2_+ - Q^2)
\nonumber\\
C^{l}_{2,+}&=&S^2 (2 Q^2_+ M^2 - Q^2_+ Q^2/2 - 2 M^2 Q^2 + Q^4)
\nonumber\\
C^{l}_{2,-}&=&S^2 Q^4/2
\nonumber
\end{eqnarray}

\end{document}